\documentclass[%
 aip,
 amsmath,amssymb,
 reprint,%
]{revtex4-1}

\usepackage{graphicx}
\usepackage{dcolumn}
\usepackage{bm}

\usepackage[utf8]{inputenc}
\usepackage[T1]{fontenc}
\usepackage{mathptmx}
\usepackage{etoolbox}

\makeatletter
\def\@email#1#2{%
 \endgroup
 \patchcmd{\titleblock@produce}
  {\frontmatter@RRAPformat}
  {\frontmatter@RRAPformat{\produce@RRAP{*#1\href{mailto:#2}{#2}}}\frontmatter@RRAPformat}
  {}{}
}%
\makeatother
\begin{document}

\preprint{AIP/123-QED}

\title[]{Preparation and Characterization of High Quality Bi$_{1-x}$Sb$_{x}$ Thin Films: A Sputtering Deposition Approach}

\author{de Almeida G. G.}
\author{de Andrade A. M. H.}%
\author{Tumelero, M. A.}
    \email{matumelero@if.ufrgs.br.}
\affiliation{Physics Institute, Federal University of Rio Grande do Sul, Av.
Bento Gon¸calves 9500, Porto Alegre, 91501970, Rio Grande do Sul, Brazil.
}%

\date{\today}

\begin{abstract}
The Bi$_{1-x}$Sb$_{x}$ was the first 3D topological insulator found in nature. It presents a complex electronic structure with topological to trivial transition and a semimetallic to semiconductor transition, both achieved by changing the x fraction of Sb. The complex nature of this system may lead to several electronic and topological phases in matter, making it a promising quantum material. Here, we focused on preparing very high-quality thin films samples of Bi$_{1-x}$Sb$_{x}$ with varying fraction of x using the Co-deposition Magnetron Sputtering technique. Our results demonstrate that high-quality samples, with compact and uniform morphology, presenting a preferential direction of growth, can be obtained over SiO$_{2}$ substrate. Our findings suggest a dependence between the thin films crystalline texture and the composition of the samples, as well as the deposition temperature.
\end{abstract}

\maketitle

%

\section{Introduction}

The Bi$_{1-x}$Sb$_{x}$ have been extensively studied in physics and material science, mainly due to its thermoelectric properties \cite{rogacheva2013electronic}. One interesting phenomenon in this system is the semimetal-semiconductor phase transitions, which occurs at x concentration of about 0.07 and 0.22 \cite{flores2020spark,colloquium-topological}. In between these x values, the system behaves as a semiconductor, while outside this range, the electronic structure is typical of semimetallic system \cite{kraak1978}. The narrow bandgap and the high electronic mobility make it an attractive material for application in thermoelectric generators and spintronics \cite{PRL-spinhall,rongione2023}. Reported values for Seebeck coefficient and electron mobility are as high as $10^{5}$ cm$^{2}$/V.s \cite{dong-xia2013} and -150 $\mu$V/K \cite{gunes2017}, respectively. 

The interest in this system has increased significantly since 2008, with the discovery of 3D topological insulators. D. Hsieh, et al \cite{hsieh2008}. Studies have shown that a topological insulator phase can be found at around x of 0.10. At x = 0.03 a band inversion take place leading to a topological phase transition, while at 0.07 the material undergoes a transition into a topological insulator phase presenting helical electronic states at the surface \cite{colloquium-topological}. 

The electronic structure of Bi$_{1-x}$Sb$_{x}$ is complex, displaying 5 surface bands and narrow band gap, making it challenging to experimentally access the electronic properties of this system \cite{hsieh2008}. Another hindrance to understanding and applying this material is the difficulty in preparing suitable samples, as the electronic properties are highly sensitive to the amount of Sb and its crystallinity.

The Bi$_{1-x}$Sb$_{x}$ has been prepared in both single-crystals \cite{sultana2019flux,kraak1978} and polycrystalline \cite{malik2012sb,flores2020spark} form. For thin films, the most common technique for growing Bi$_{1-x}$Sb$_{x}$ is Molecular Beam Epitaxy (MBE) \cite{cho1999, rongione2023, ueda2020fabrication}, which yields high-quality samples. However, MBE is both costly and time consuming, limiting the number of studies and the potential for technological applications. Reports using magnetron sputtering have indicated the possibility of growing high-quality Bi$_{1-x}$Sb$_{x}$  samples, comparable to those produced by MBE \cite{tuofan2020, rochford2015controlling,zhendong2020}. Rochford et al. \cite{rochford2015controlling}, shown that the thermal treatment of sputter-deposited Bi$_{1-x}$Sb$_{x}$ can enhance the crystalline texture, and the use of a capping layer can reduce the Sb segregation and oxidation. Sputter deposited thin films have also been used to investigate spin-Hall effect in Bi$_{1-x}$Sb$_{x}$ \cite{zhendong2020}. 

In order to better understand the growth process of Bi$_{1-x}$Sb$_{x}$ using sputtering methods and to identify the deposition conditions that yield Bi$_{1-x}$Sb$_{x}$ samples suitable for both fundamental scientific research and technological application, here, we investigate the magnetron sputtering growth of Bi$_{1-x}$Sb$_{x}$ thin films. We employed a single-step deposition method by using heated substrates. Our findings suggest an easy method to control the stoichometry Bi$_{1-x}$Sb$_{x}$ by adjusting the relative deposition power in co-deposition process. Additionally, our results indicate that both the preferential growth direction and the crystallinity of the samples depend on the deposition temperature and the antimony concentration in Bi$_{1-x}$Sb$_{x}$.

\section{Methods}

The samples were prepared using a Magnetron Sputtering AJA Orion-8-UHV, under base pressure of 5$\times$10$^{8}$ torr. The co-deposition method was employed, with simultaneous deposition  of Bismuth (99.999 \%) and Antimony (99.999 \%). A silicon wafer coated with 100 nm of SiO$_{2}$ was used as substrate, which was kept rotating throughout the deposition process to improve sample uniformity. The sample holder temperature was maintained in the range of RT to 260 ºC to enhance samples crystallinity. The argon working pressure was set to 2 mtorr, and the target power was adjusted within the range of 10 and 25 W of RF power. The XRD analysis was performed using a Bruker D8 Advanced diffractometer in a $\theta-2\theta$ configuration. The sample thickness was determined using a quartz microbalance inside the deposition chamber prior to the deposition, and after confirmed by using X-ray reflectometry and Rutherford backscattering spectroscopy (RBS). RBS was also used to determine the Bi and Sb concentration profile in the thin films. Electrical properties was measured in a custom setup using a Cryomagnetic Criocoller CMAG9 criostat. For resistivity and Hall effect measurements we have applied the standard 4-probe configuration in a Hall bar geometry.

\section{Results}

\begin{figure}[h]%
\centering
\includegraphics[width=0.5\textwidth]{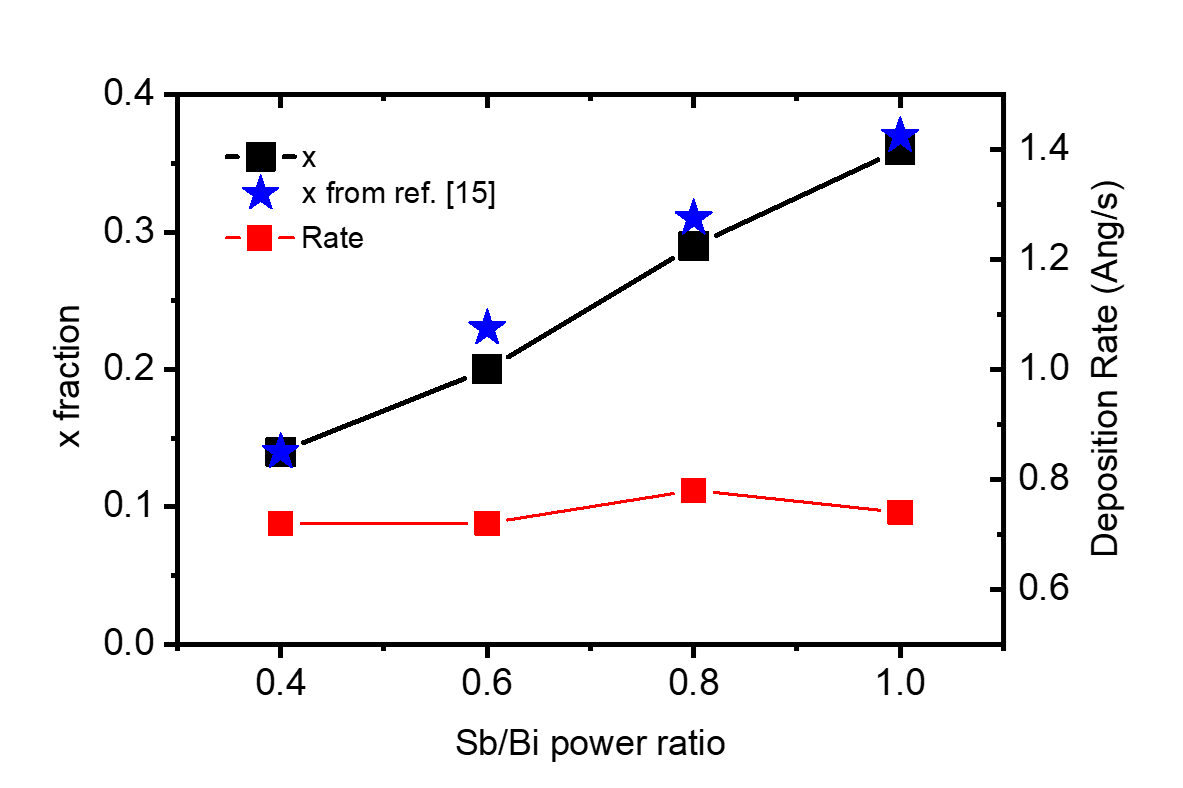}
\caption{The x fraction on Bi$_{1-x}$Sb$_{x}$ thin films as a function of the Sb/Bi sputter power ratio. The data in star symbols is reproduced from ref. \cite{rochford2015controlling}.}\label{fig1}
\end{figure}

In Figure \ref{fig1} is presented the Bi$_{1-x}$Sb$_{x}$ x fraction obtained from RBS analysis for different Sb/Bi power ratios used in the co-deposition process with the substrate temperature set at 260 ºC. It is indicating that this parameter can be used for a precise control of the x-fraction. Intriguingly, the deposition rate appears to be independent of the overall deposition power (Bi + Sb power), at least within the range used in this work. As shown on the right-hand axis of the Figure \ref{fig1} (in red), the deposition rate is approximately 0.75 A/s. All the samples prepared in this investigation were handle to have a thickness of about 120 nm thickness. Additionally, the blue stars represent data from ref. \cite{rochford2015controlling}, where a similar relationship between the x-fraction dependence on the Sb/Bi power ratio was reported.

\begin{figure}[h]%
\centering
\includegraphics[width=0.5\textwidth]{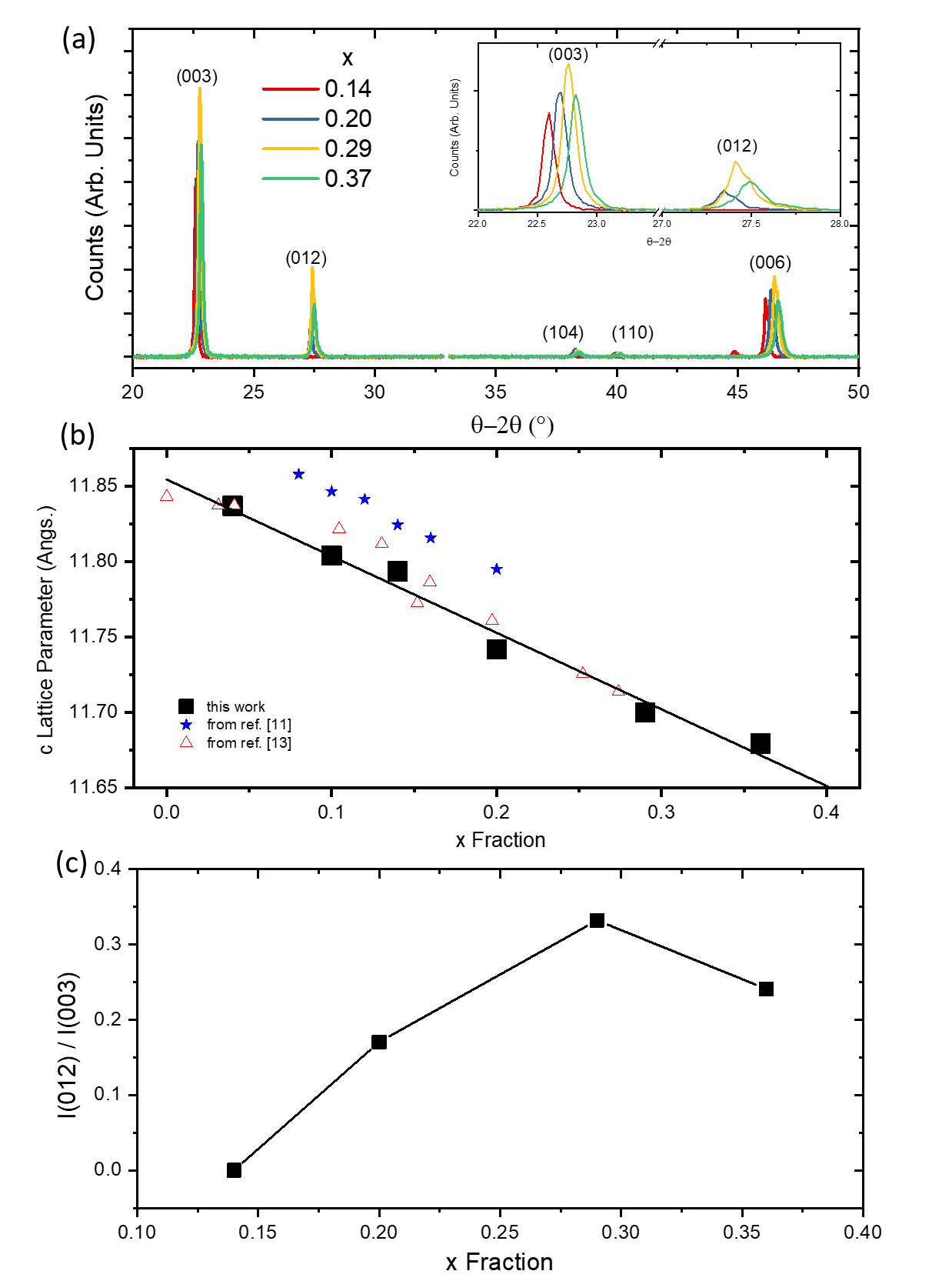}
\caption{In (a) the XRD pattern for samples prepared with distinct x fraction. The inset in (a) displays a zoom view at the peak (003). In (b) is presented the lattice parameter c as a function x. (c) Intensity ratio between peaks (012) and (003) at samples with distinct x.}\label{fig2}
\end{figure}

Figure \ref{fig2}(a) displays the XRD results for samples with x fractions of 0.14, 0.20, 0.29, and 0.37. Peaks corresponding to the (003), (012), (104), (110), (006), and (202) planes can be observed. The pattern is consistent with the R-3m phase of Bi and Sb, and the presence of a single sharp peak indicates the successful deposition of a well-formed Bi-Sb random alloy. Figure \ref{fig2}(b) shows the c-lattice parameter as a function of the x fraction. As the x fraction increases, the lattice parameter decreases, which is typically associated with the substitution of Sb atoms into Bi lattice sites. The red triangles represent data from ref. \cite{ueda2020fabrication}, obtained from Bi${1-x}$Sb${x}$ thin films deposited via MBE, while the blue stars represent data from ref. \cite{malik2012sb}. The lattice parameters align well with previously reported values for high-quality samples.

The pattern shown in Figure \ref{fig2}(a) differs significantly from what is typically expected in powder diffraction, indicating the presence of crystalline texture with a preferential growth direction along (003) planes. Figure \ref{fig2}(c) presents the intensity ratio of the (012) and (003) peaks as a function of the x fraction. The texture appears to be dependent on the x fraction, with an increase in x resulting in a relative decrease in the (003) peak intensity compared to (012), suggesting a reduction in crystalline texture. It is noteworthy that all samples were deposited under identical conditions and have the same thickness. Interestingly, for an x fraction of approximately 0.14, only the (003) and (006) peaks are observed, indicating that these samples exhibit a fully textured structure.

\begin{figure}[h]%
\centering
\includegraphics[width=0.5\textwidth]{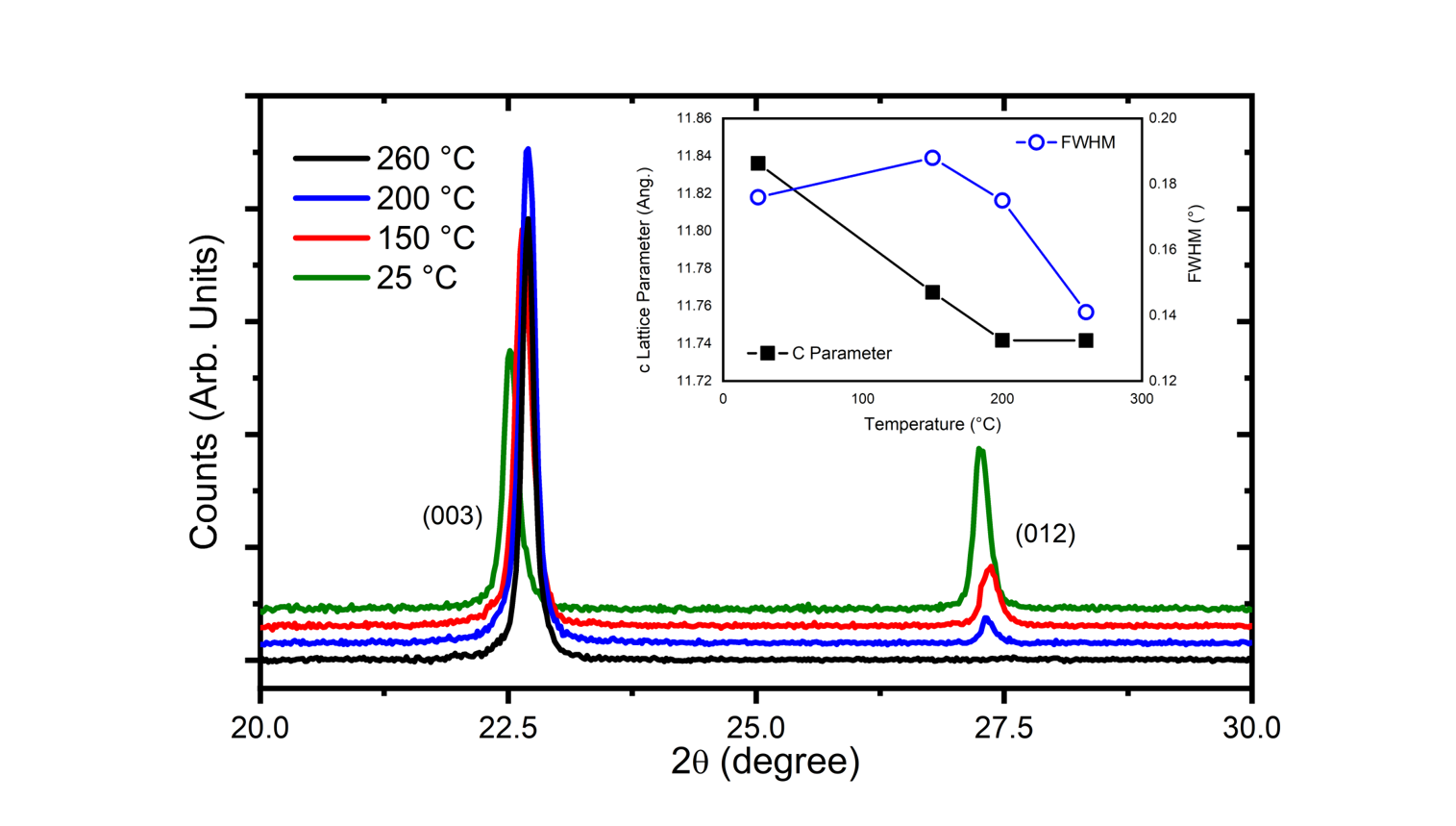}
\caption{XRD pattern for samples deposited with Sb/Bi power ratio of 0.6 at distinct substrate temperatures. In the inset, the lattice parameter and the FWHM of the peak from planes (003).}\label{fig3}
\end{figure}

Figure \ref{fig3} presents the XRD pattern for samples deposited with a constant Sb/Bi power ratio of 0.6 and prepared at different deposition temperatures. It can be observed that as the deposition temperature is reduced to 200ºC and below, the intensity of the (012) peak increases relative to the (003) peak, indicating that fully [003]-textured samples can only be obtained at deposition temperatures near the melting point of Bi. The inset of Figure \ref{fig3} shows the Full Width at Half Maximum (FWHM) of the (003) peak for the different deposition temperatures. A noticeable increase in the FWHM occurs as the deposition temperature decreases, suggesting that deposition temperatures close to Bi melting point also improve the crystallinity of the samples. At 260°C, the FWHM is approximately 0.14, which is smaller than values previously reported for MBE deposition \cite{cho1999-2}, confirming the high quality of the samples. Additionally, the inset also shows the c-lattice parameter as a function of deposition temperature. As the temperature decreases from 260°C to near room temperature (RT), the c parameter increases from 11.77 Å to approximately 11.81 Å.

The electrical properties of the samples were investigated through $\rho \times$ T measurements. The results for samples with varying Sb concentrations are shown in Figure \ref{fig4}. Samples with x = 0.14 and 0.20 exhibit a negative temperature coefficient, with resistivity saturating at low temperatures, consistent with the semiconductor behavior expected for these Sb concentrations \cite{cho1999, flores2020spark}.
In contrast, samples with x = 0.29 and 0.37 display a more complex behavior at higher temperatures, where resistivity begins to increase above 225 K. This behavior may be attributed to the complex electronic structure of the semimetal phase of Bi${1-x}$Sb${x}$ for x $>$ 0.22 \cite{kraak1978}. At room temperature (RT), the resistivity for the sample with x = 0.14 is approximately 420 $\mu\Omega \cdot \text{cm}$, which is comparable to values found in ref. \cite{malik2012sb, flores2020spark, tuofan2020} for polycrystalline samples. However, it is higher than what is typically reported for single crystals and MBE-prepared samples, as in ref. \cite{nishide2010, huang2024}. For the sample with x = 0.20, the RT resistivity increases to about 760 $\mu\Omega \cdot \text{cm}$, while for x = 0.29, it reaches approximately 890 $\mu\Omega \cdot \text{cm}$. Interestingly, further increases in the Sb concentration (x = 0.37) lead to a decrease in resistivity, with values around 580 $\mu\Omega \cdot \text{cm}$.

Additionally, in samples with x = 0.14 and 0.20 — both of which fall within the semiconductor region of the phase diagram — the resistivity exhibits saturation at low temperatures. Specifically, for x = 0.14, the resistivity decreases before reaching saturation, as shown in Figure \ref{fig4}, resulting in a maximum resistivity value at approximately 80 K. This maximum in the $\rho$ vs. T relationship has been reported by other research groups, including those in ref. \cite{ueda2020fabrication, malik2012sb, flores2020spark, tuofan2020}, all of which focused on polycrystalline samples.

\begin{figure}[h]%
\centering
\includegraphics[width=0.48\textwidth]{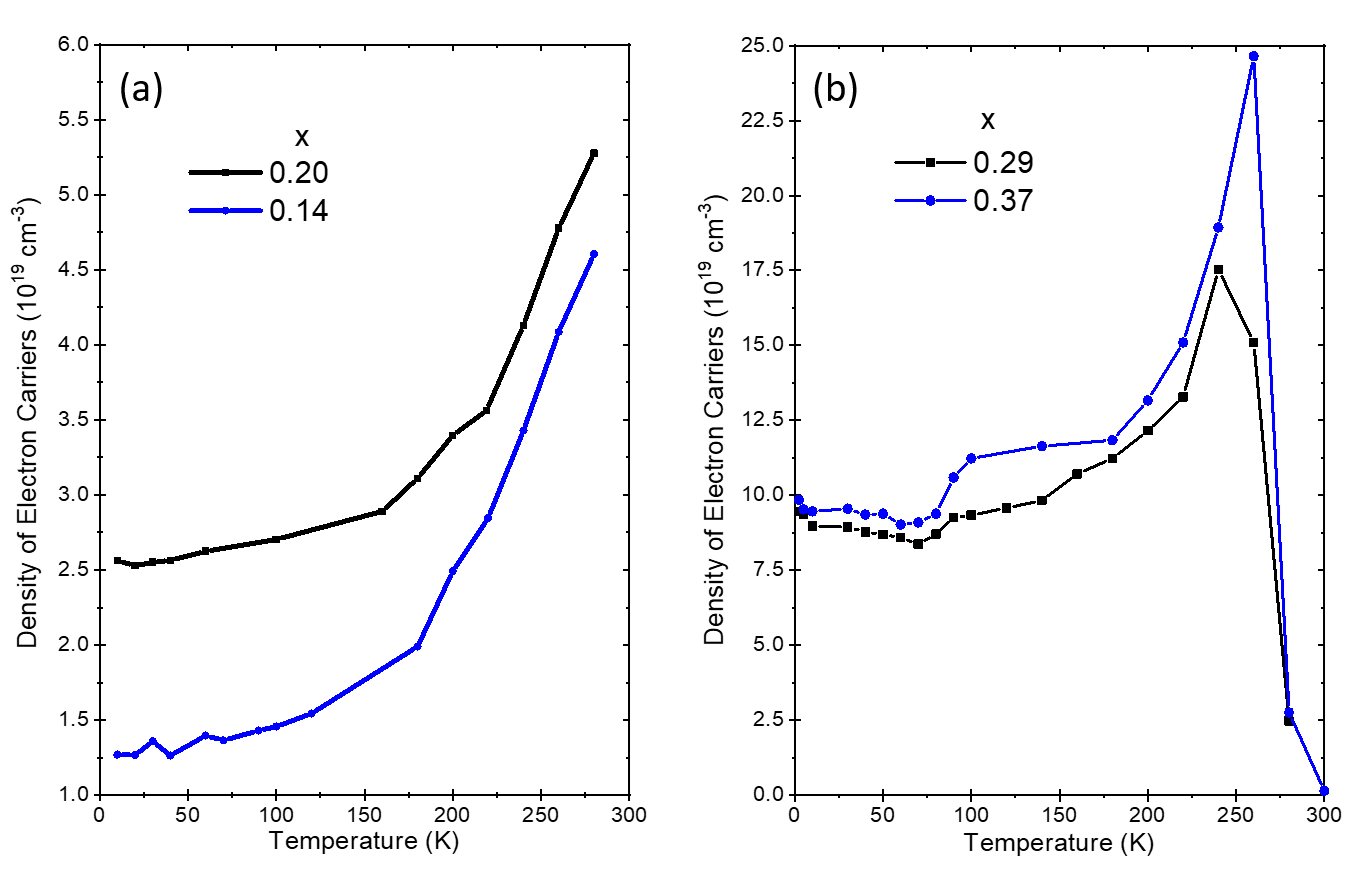}
\caption{Density of carriers vs. temperature, obtained from Hall effect measurements, in (a) for samples with x =0.14 and 0.20 and in (b) for x=0.29 and 0.37.}\label{fig5}
\end{figure}

The charge carrier densities for the various samples are illustrated in Figure \ref{fig5}. The carrier density was determined by measuring the Hall coefficient at different temperatures. The data are presented in two separate graphs: one for samples with x = 0.14 and 0.20 (semiconductor) and another for samples with x = 0.29 and 0.36 (semimetal). This division allows for an expanded view of the data scales. In both cases, the carriers are electrons. In Figure \ref{fig5}(a), we observe that the carrier density ranges from 1 to 6 $\times 10^{19}$ cm$^{-3}$; meanwhile, in Figure \ref{fig5}(b), the carrier density is approximately ten fold higher. This significant difference in carrier density values between the x = 0.14 and 0.20 samples compared to the 0.29 and 0.36 samples further confirms the distinct semiconducting and semimetallic phases, respectively. For all samples, the charge carrier density increases with rising temperature, suggesting that even in samples exhibiting semimetallic behavior, there are likely pseudogaps on the Fermi surface that become populated as the temperature increases. Additionally, Figure \ref{fig5}(a) shows that the carrier density for the sample with x = 0.14 is lower than that of the sample with x = 0.20, which is consistent with the presence of a larger bandgap in the former.

To compute the electronic bandgap, assuming that the increase of carrier density is essentially due to the promotion of electrons from valence band through the bandgap, we used the equation for the thermal activated transport, as presented below:   

\begin{equation}
   n(T) \propto n_{0}  exp\left(\frac{E_{G}}{2k_{B}T}\right),
   \label{eqEgap}
\end{equation}

where $k_{B}$ is the Boltzmann constant, $E_{A}$ is the activation energy of the carriers, which could be assumed to be the bandgap. The energy values found were 20.7, 11.7, 3.5 and 2.8 meV for the samples with X = 0.14, 0.20, 0.29 and 0.37, respectively. The value for the sample with x = 0.14 is small than previous reported in ref. \cite{cho1999,malik2012sb,flores2020spark}, but close to the values reported in ref. \cite{LENOIR1996}. 

\begin{figure*}[ht]%
\centering
\includegraphics[width=1\textwidth]{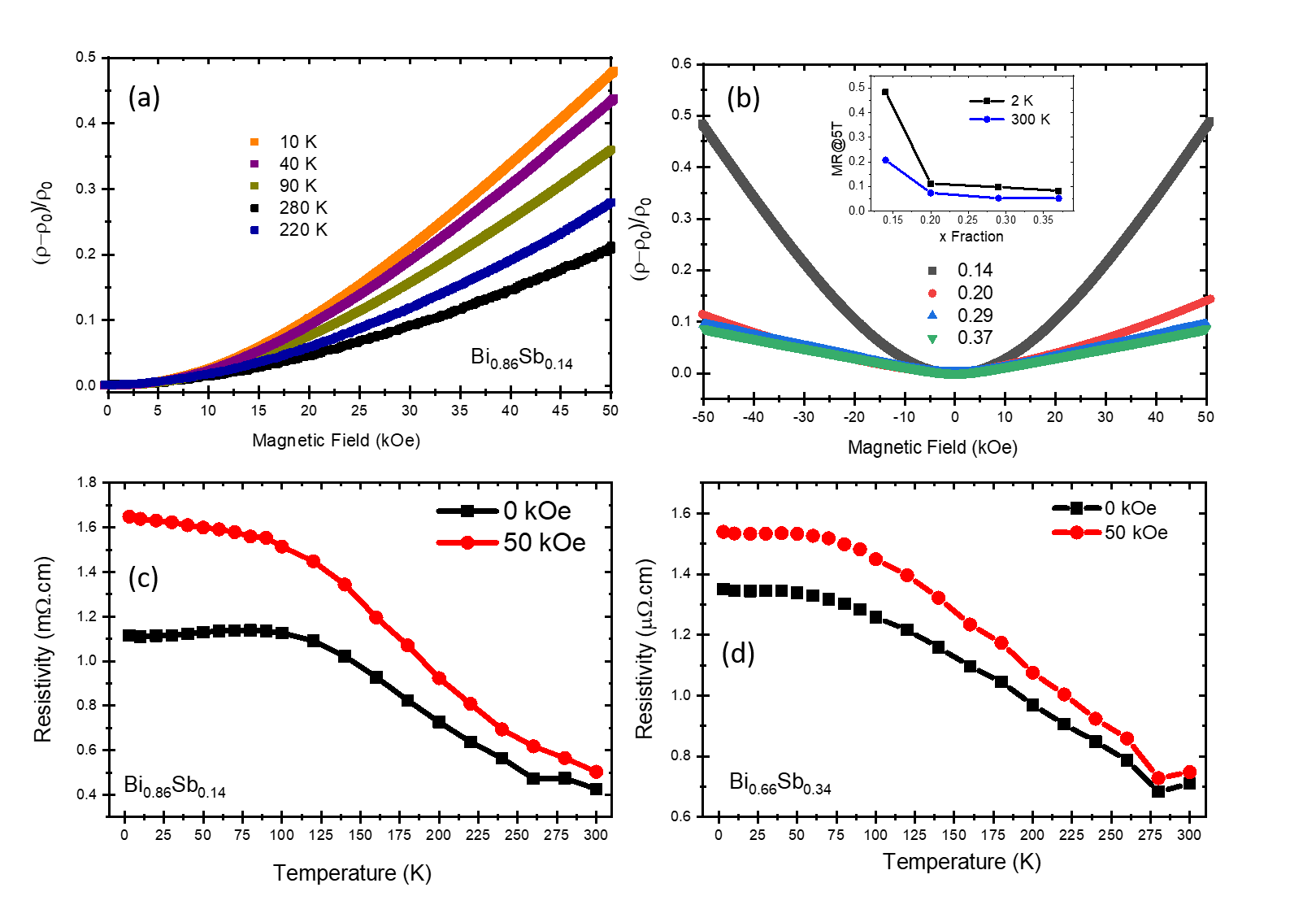}
\caption{Magnetic field dependence of electrical resistivity (out-of-plane field). In (a) for samples with x=0.14 at distinct temperatures, (b) at 2K for distinct samples. The inset shows the MR vs. x at 2K and 300K. (c) The resisitity vs. temperature under 0 and 50 KOe for the samples with x = 0.14 and (d) for the samples with x = 0.20.}\label{fig6}
\end{figure*}

The magnetoelectric properties of Bi${1-x}$Sb${x}$ were investigated through magnetoresistance (MR) measurements. Figure \ref{fig6}(a) presents the MR curves for the sample with x = 0.14 at various temperatures. At low magnetic fields, the electrical resistance increases with the magnetic field in a square-law fashion, which is characteristic of ordinary cyclotron scattering of the charge carriers. The observed increase in MR with decreasing temperature is attributed to enhanced carrier mobility at low temperature. At high magnetic fields and low temperatures, the magnetoresistance exhibits linear behavior. This type of behavior has been reported in several semimetallic, topological, and trivial systems and may be associated with energy band scattering effects, as well as the influence of disorder \cite{LMR-doi:10.1021/acs.nanolett.1c01647, LMR-doi:10.1063/1.5127570, LMR-Hu2008, LMR-PhysRevB.102.115145}. The values of MR at low temperatures, approximately 50\% at 10 K, are significantly smaller compared to those observed in single crystals with x = 0.03 \cite{maurya2020magnetotransport}. No comparable data in the literature is available for samples with x values similar to those reported here. It is important to note that an increase in Sb content is expected to drastically reduce electron mobility, thereby decreasing the magnetoresistance of the samples. This is consistent with the data presented in Figure \ref{fig6}(b), which shows MR at 2 K for samples with x = 0.14, 0.20, 0.29, and 0.36. The inset of Figure \ref{fig6}(b) indicates a reduction in magnetoresistance with increasing x ratio at both 2 K and 300 K. Figures \ref{fig6}(c) and (d) display the $\rho$ vs. T curves for the semiconducting samples with x = 0.14 and x = 0.20, both with and without a magnetic field (5 T). For the sample with x = 0.14, the application of a magnetic field eliminates the resistivity maximum at approximately 100 K. This behavior may be related to topological surface effects occurring within this range of x.

\section{Conclusions}

In this work, we prepared Bi${1-x}$Sb${x}$ samples with varying Sb concentrations using the co-deposition magnetron sputtering technique, demonstrating that substrate temperature is a major factor influencing grain growth and film quality. By depositing at 260°C, we obtained a fully textured sample with a preferential orientation along (003). The texture is highly dependent on both the substrate temperature during deposition and the Sb concentration. The electrical resistivity as a function of temperature indicates that samples with x = 0.14 and 0.20 exhibit semiconductor behavior, while the samples with x = 0.29 and 0.36 are typical semimetals. The carrier concentration suggests the presence of a bandgap of about 20 meV for the samples with x = 0.14, which decreases with increasing Sb content. At low temperatures and high magnetic fields, we observed a linear dependence of resistivity on the magnetic field, which could be related to the topological nature of this compound.

\begin{acknowledgments}
The authors would like to acknowledge the following grants: CNPQ 402915/2021-6, CNPQ 313809/2023-2, FAPERGS. The authors would like to acknowledge the Nanoconformation Laboratory (LCI) and Ion Implantation Laboratory (LII) for the XRD and RBS facilities, respectively. The authors would like to thank the Prof. Paulo Pureur for the support to this work.
\end{acknowledgments}

\nocite{*}
\bibliography{aipsamp}

\begin{figure*}[ht]%
\centering
\includegraphics[width=1\textwidth]{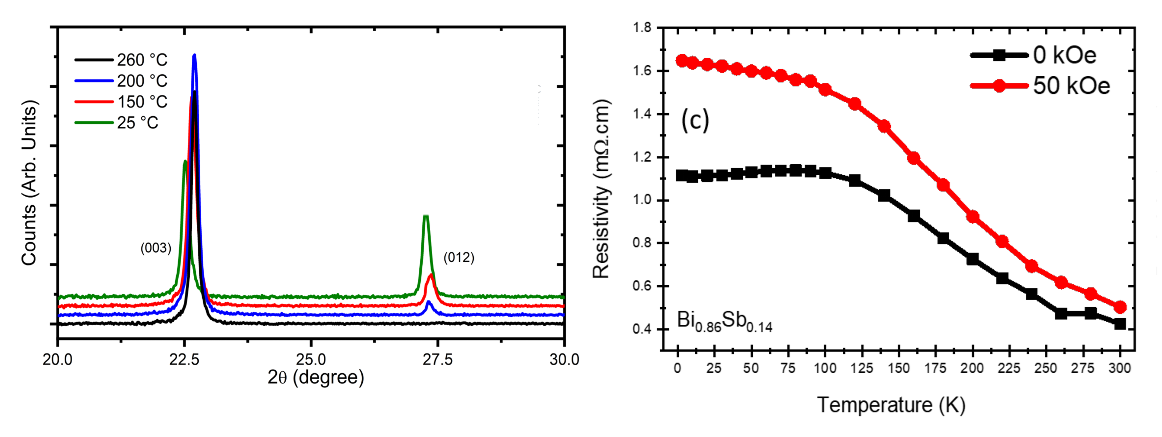}
\caption{Graphical Abstract}\label{fig6}
\end{figure*}

\end{document}